# More on solar-like oscillations in $\eta$ Boo

T.R. Bedding

*School of Physics, University of Sydney 2006, Australia*

H. Kjeldsen

*Institute of Physics and Astronomy, Aarhus University, DK-8000 Aarhus C, Denmark*

**Abstract.** We recently reported strong evidence for solar-like oscillations in the G0 IV star $\eta$ Boo (Kjeldsen et al. 1995, AJ **109,** 1313). We measured small temperature fluctuations produced by oscillations through their effect on the equivalent widths of the Balmer lines. Here we address several issues that were raised at this conference.

**Removing the high-pass filter** To improve the estimates of equivalent-width measurements, the time series was decorrelated against external parameters. This is a well-established procedure (Brown et al. 1991) in which one computes a correction to each data point. The result for $\eta$ Boo was a reduction in the rms scatter of the time series by a factor of 1.9. However, in order to calculate the decorrelation corrections, it was necessary to apply a high-pass filter to remove variations on long timescales. The power spectrum in Kjeldsen et al. (1995) clearly shows the effects of this filter.

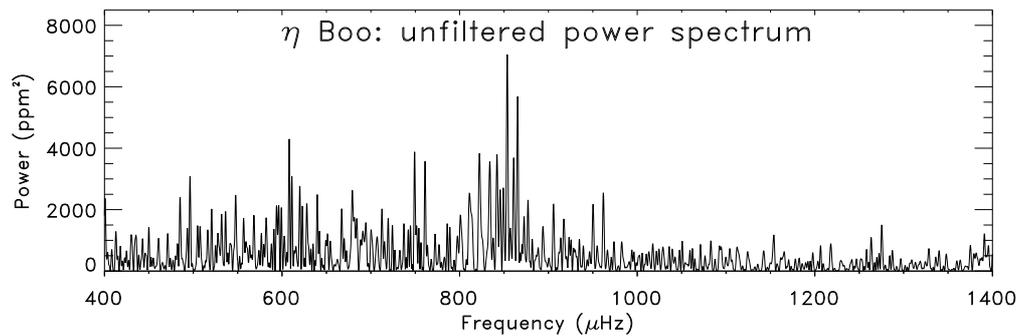

It would clearly be desirable to generate a power spectrum that has not been high-pass filtered. This is possible, since the filtering is only necessary while the decorrelation is being performed. Once the decorrelation corrections have been calculated, they can then be applied to the original *unfiltered* time series. The resulting power spectrum for our $\eta$ Boo data is shown in the figure. Comparison with Fig. 1 in Kjeldsen at al. (1995) shows an increase in noise at low frequencies, as expected from a $1/f$ noise source. Apart from that, the two are very similar and the oscillation signature is still clearly visible as a hump of excess power in the frequency range 750–950 $\mu$Hz.

**Solar oscillations** During the daylight hours we obtained spectra of the Sun by observing scattered light from the sky. As described in Kjeldsen et al. (1995), we were able to detect the solar oscillations in a folded power spectrum. At the time, we were puzzled by the detection of the $l = 3$ modes in the Sun, given that the observations did not resolve the solar disk. We now understand this. Model atmosphere calculations show that the Balmer lines are much weaker towards the solar limb, so our measurements of equivalent width were weighted to the central portions of the disk (Bedding et al., in preparation).

**Future observations of $\alpha$ Cen A** We will use the equivalent-width method to search for oscillations in $\alpha$ Cen A with the AAT 3.9 m and ESO 3.6 m over six nights in April 1995. The figure shows simulations of the power spectrum expected from this type of observation. Labels indicate modes with $l = 0$ and 1; modes with $l = 2$ are shown by dashed lines. The insets show the window functions. Input frequencies were from Edmonds et al. (1992). The input amplitudes had a solar-like envelope, randomized about an average peak value of 8 ppm (which is 1.3 times that of the Sun; see Kjeldsen & Bedding 1995). We assumed photon-noise limited observations of the H$\alpha$ line, with a detected photon flux of $4 \times 10^6$ s$^{-1}$Å$^{-1}$ for 9 hours per night for each telescope.

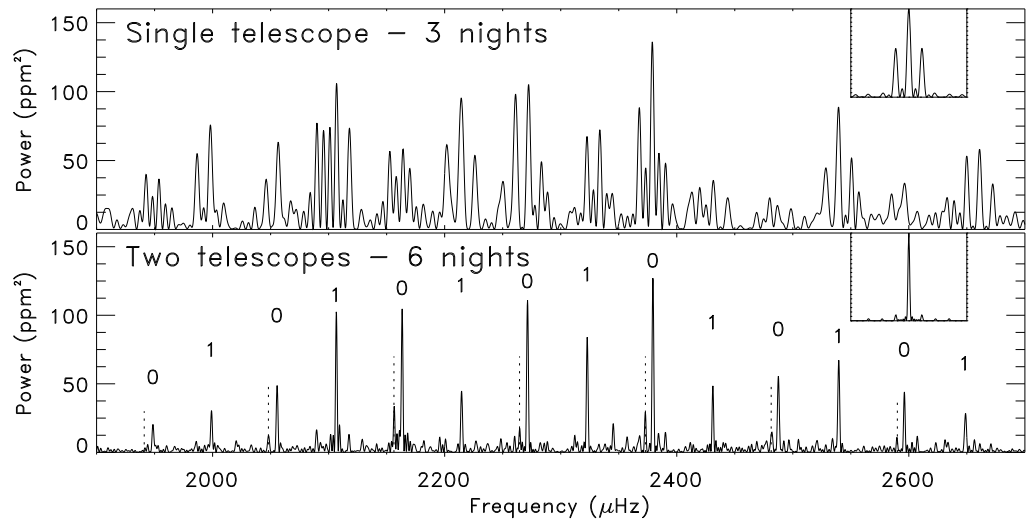